\documentclass[aps, 10pt, prb, showkeys, twocolumn, superscriptaddress]{revtex4-2}
\usepackage{graphicx, amssymb, amsmath, amsfonts, enumitem}
\usepackage[utf8]{inputenc}
\usepackage[normalem]{ulem}
\usepackage[colorlinks=true, linkcolor=blue, citecolor=blue, urlcolor=blue]{hyperref}
\usepackage{url, soul, ulem, textgreek, bm, comment}
\usepackage[dvipsnames]{xcolor}
\usepackage{appendix}
\usepackage{kotex}
\usepackage{float}
\usepackage{orcidlink}
\usepackage{booktabs}
\usepackage{balance}

\providecommand{\up}{\uparrow}
\providecommand{\dn}{\downarrow}

\newcommand{\apctpadd}{Asia Pacific Center for Theoretical Physics, Pohang, Gyeongbuk, 37673, Republic of Korea}
\newcommand{\postechadd}{Department of Physics, Pohang University of Science and Technology, Pohang, Gyeongbuk 37673, Republic of Korea}

\begin{document}

\title{Equal-spin pairing at finite Cooper-pair momentum in $p$-wave magnets}

\author{Jaeyong Lee\orcidlink{0009-0007-1827-3866}}
\affiliation{\postechadd}

\author{Kyoung-Min Kim\orcidlink{0000-0003-2468-3152}}
\email{kyoungmin.kim@apctp.org}
\affiliation{\apctpadd}
\affiliation{\postechadd}

\date{\today}

\begin{abstract}
The realization of Fulde--Ferrell (FF) superconducting states has long been pursued but remains challenging because of the large Zeeman splitting required. Here, we demonstrate that $p$-wave magnets can host equal-spin finite-pair-momentum states with $p$-wave symmetry, termed FF states, at arbitrarily weak fields. Our Bogoliubov--de Gennes analysis shows that, owing to the displacement of the Fermi surface, two equal-spin electrons in the FF states pair at the same energy, just as in the opposite-spin zero-pair-momentum states of the same gap symmetry. At zero field, the FF states are therefore exactly degenerate with these zero-pair-momentum counterparts. An arbitrarily weak Zeeman field lifts this degeneracy and makes the FF states the superconducting ground state, since they remain stable owing to their equal-spin nature, whereas their counterparts are suppressed by the formation of Bogoliubov Fermi surfaces. These findings identify the fermiology of $p$-wave magnets as a promising route toward long-sought finite-momentum superconductivity.
\end{abstract}


\maketitle

\section{Introduction}

Finite-momentum Cooper pairing---in which the Cooper pair carries a nonzero center-of-mass momentum---has been a longstanding pursuit in the study of unconventional superconductivity~\cite{Casalbuoni2004, Agterberg2020}. The paradigmatic example is the Fulde--Ferrell--Larkin--Ovchinnikov state, in which pairing of two electrons having nonzero net momentum is stabilized by a Zeeman-split Fermi surface (FS), allowing electrons of opposite spin to pair while conserving energy~\cite{Fulde1964, Larkin1965}. To realize such finite-momentum pairing, researchers have conventionally considered an external magnetic field or a ferromagnetic exchange field~\cite{Demler1997, Ryazanov2001, Kontos2002}, both of which are accompanied by a finite net magnetization that limits their practical utility, for instance by inducing vortices via orbital coupling to the applied field~\cite{Matsuda2007}. Field-free routes to finite-momentum pairing have also been explored, most notably as a pair-density wave (PDW) intertwined with stripe order in doped Mott insulators~\cite{Berg2009, Fradkin2015}, or as Amperean pairing mediated by gauge-field fluctuations~\cite{Lee2014}; however, these mechanisms rely on strong correlations or proximity to a stripe-ordered state, rather than providing a generic route accessible in a clean, weakly correlated metal.

Recently, unconventional antiferromagnets have emerged as a novel platform for realizing finite-momentum pairing states without requiring a net magnetization~\cite{Mazin2025, Fukaya_2025}, potentially circumventing the constraints faced by the mechanisms above. This class of collinear magnets combines zero net magnetization with momentum-dependent spin splitting, a combination that has recently been experimentally confirmed in several materials~\cite{Lee2024,Krempasky2024,Ding2024,Reimers2024,Yang2025,Sandra2025,Zhang2025,Jiang2025}. Such spin splitting can suppress the conventional $s$-wave spin-singlet channel, opening a pathway toward finite-momentum superconducting states. On the weak-coupling side, a series of theoretical studies established that the spin splitting alone can stabilize finite-momentum Cooper pairing, whether induced by proximity~\cite{Zhang2024}, intrinsic attractive interactions~\cite{PhysRevB.110.L060508,Sim2025}, or an exchange-field-generated Bogoliubov FS~\cite{Hong2025}, with follow-up works~\cite{Hu2025,Liu2026}. Beyond this mean-field picture, interaction-driven approaches have also identified finite-momentum pairing, arising from spin fluctuations in a weakly correlated Hubbard model~\cite{Parthenios2025} and from strong correlations in a doped Mott insulator~\cite{Chen2026}. Recently identified odd-parity magnets---$p$-wave magnets~\cite{Yamada2025, Hellenes2023, Sim2026} and $f$-wave magnets~\cite{Zeng2026, Zhuang2026}---are similarly expected to offer a possible route toward finite-momentum pairing via the momentum-dependent spin splitting induced by the exchange field. Despite this potential, existing studies have largely focused on superconducting phenomena within the scope of zero-momentum Cooper pairing~\cite{PhysRevB.111.174519, Sun2025, Sukhachov2025, Khodas2026, Kim2026a, Pal2026}, and finite-momentum pairing states in odd-parity magnets remain elusive.

\begin{figure}[t!]
    \centering
    \includegraphics[width=\linewidth]{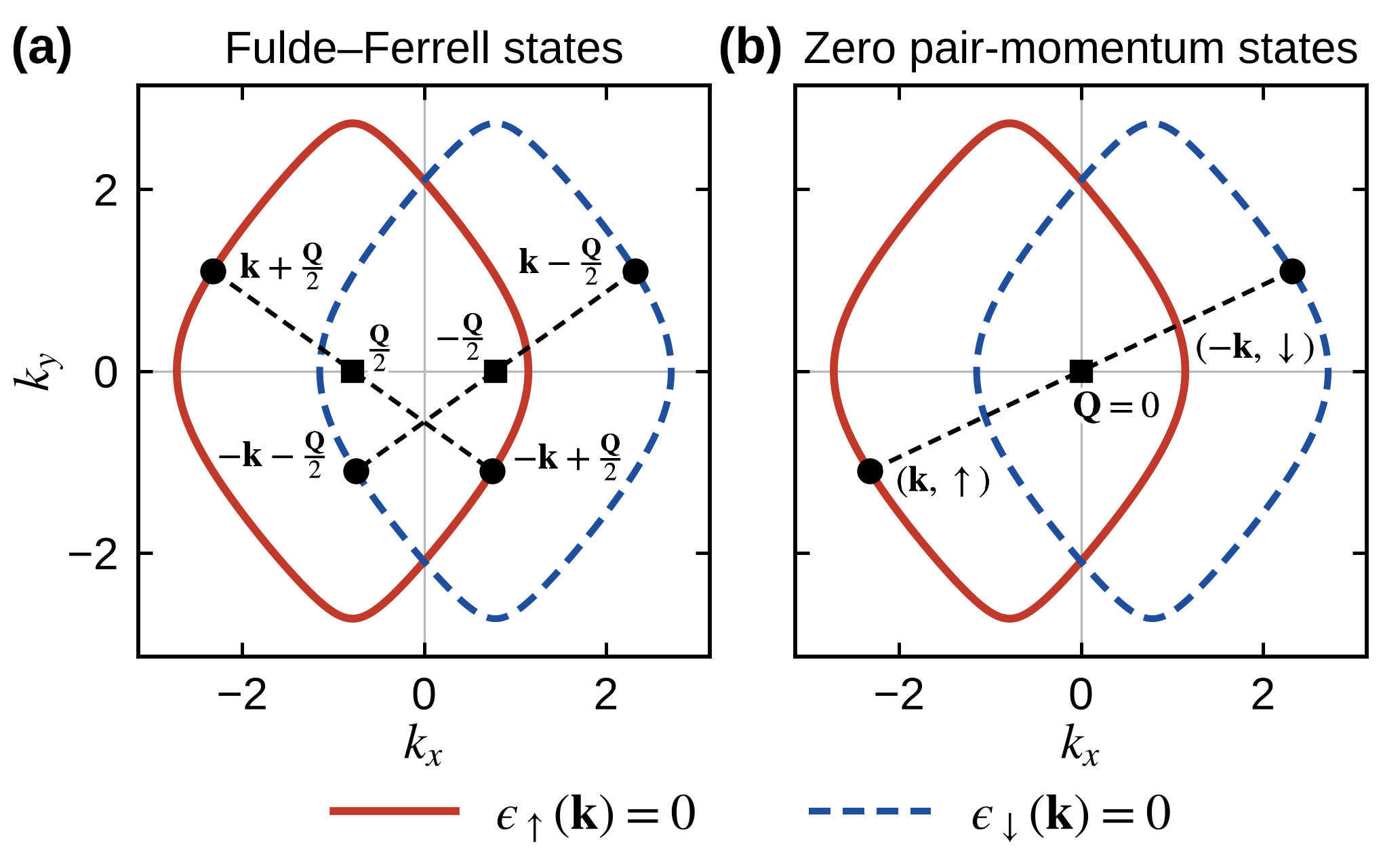}
    \caption{\textbf{Schematic illustration of the pairing channels.} (a) Equal-spin finite-momentum pairing in the Fulde--Ferrell (FF) states. The red and blue lines represent the up- and down-spin FSs, respectively, each shifted by $+\frac{\bm{Q}}{2}$ and $-\frac{\bm{Q}}{2}$, with $\bm{Q}=Q\hat{\bm{k}}_x$ and $Q=-2\tan^{-1}\left(\frac{J}{t}\right)$. The vectors $\bm{k}+\frac{\bm{Q}}{2}$ and $-\bm{k}+\frac{\bm{Q}}{2}$ ($\bm{k}-\frac{\bm{Q}}{2}$ and $-\bm{k}-\frac{\bm{Q}}{2}$) denote the momenta of the two paired spin-up (spin-down) electrons, with the pair momentum $\bm{Q}$ ($-\bm{Q}$). (b) Inter-spin zero-momentum pairing, shown on the same FSs as in (a). The parameters used are $J=1$ and $\mu=-1$.}
\label{fig1}
\end{figure}

In this work, we establish a finite-momentum equal-spin pairing instability intrinsic to a two-dimensional $p$-wave magnet on a square lattice. There, the $p_x$-wave exchange field displaces the two spin-split FSs in opposite directions along $k_x$~\cite{Kim2026a, Khodas2026}. We introduce equal-spin, odd-parity finite-momentum pairing states whose pairing momentum is set by this displacement, hereafter referred to as ``Fulde--Ferrell (FF) states,'' although the original FF state denotes pairing between electrons of opposite spin with even-parity form factors~\cite{Fulde1964}. Our Bogoliubov--de Gennes (BdG) analysis shows that two electrons of the same spin are paired at the same energy at every point of the FS [Fig.~\ref{fig1}(a)], which maximizes the pairing amplitude and makes the FF states degenerate with the relevant zero-momentum pairing states [Fig.~\ref{fig1}(b)]. Applying a Zeeman field then lowers the energy of the FF states below that of the zero-momentum pairing states, stabilizing the FF states as the energetically favored superconducting phases. We finally show that the quasiparticle energy dispersions under a Zeeman field provide an experimental probe that distinguishes the FF states from the zero-momentum pairing states, which are identical at zero field.

\section{Model}

\subsection{Extended attractive Hubbard model}

We consider an effective two-band model on a square lattice for $p$-wave magnets~\cite{Kim2026a, Khodas2026},
\begin{equation} \label{eq:model}
\begin{aligned}
    \hat{H} = & \; \hat{H}_0 + \hat{H}_\mathrm{int}, \\
    \hat{H}_0 = & \; -t\sum_{\sigma} \sum_{\langle i,j\rangle} (c_{i\sigma}^\dagger c_{j\sigma} + c_{j\sigma}^\dagger c_{i\sigma}) \\
    & \; + \frac{J}{\imath} \sum_{\sigma} \sum_{i} s_{\sigma}  (c_{i\sigma}^\dagger c_{i+x\sigma} - c_{i+x\sigma}^\dagger c_{i\sigma}) \\
    & \; -  h\sum_{\sigma} \sum_{i} s_{\sigma}n_{i\sigma} - \mu \sum_{\sigma} \sum_{i} n_{i\sigma}, \\
    \hat{H}_\mathrm{int} = & \; -U\sum_{i} n_{i\uparrow}n_{i\downarrow} - V \sum_{\langle i, j\rangle} \sum_{\sigma,\sigma'} n_{i\sigma} n_{j\sigma'},
\end{aligned}
\end{equation}
where $c_{i\sigma}^\dagger$ ($c_{i\sigma}$) creates (annihilates) an electron at site $i$ with spin $\sigma$, and $n_{i\sigma} = c_{i\sigma}^\dagger c_{i\sigma}$ is the number operator; the summation over $\sigma,\sigma' \in \{\uparrow,\downarrow\}$ is implied throughout. The quadratic Hamiltonian $\hat{H}_0$ consists of nearest-neighbor hopping with amplitude $t$, a chemical potential $\mu$, an external Zeeman field $h$, and a $p$-wave exchange field $J$. The exchange term couples only sites separated along the $x$ direction, with the $+x$ and $-x$ hopping processes carrying opposite phase. The resulting spin splitting is odd in momentum and carries opposite sign for spin-up and spin-down electrons, with $s_\uparrow = +1$ and $s_\downarrow = -1$. The interaction part $\hat{H}_\mathrm{int}$ contains an on-site attractive interaction $U$ and a nearest-neighbor density-density interaction $V$. All energies are given in units of $t$, and we set $t=1$ throughout.

In momentum space, the quadratic Hamiltonian $\hat{H}_0$ is written as
\begin{equation}
    \hat{H}_0 = \sum_{\bm{k},\sigma} \epsilon_{\sigma}(\bm{k})c_{\bm{k}\sigma}^\dagger c_{\bm{k}\sigma},
\end{equation}
where the kinetic energy dispersion $\epsilon_{\sigma}(\bm{k})$ is given by
\begin{equation} \label{eq:E_ns_1}
\begin{aligned}
    \epsilon_{\sigma}(\bm{k}) = & \; -2t\cos k_x -2t\cos k_y + 2J s_{\sigma} \sin k_x \\ 
    & \; - h s_{\sigma} - \mu .
\end{aligned}
\end{equation}
This dispersion can be rewritten as
\begin{equation} \label{eq:E_ns_2}
    \epsilon_{\sigma}(\bm{k}) = -2t_1 \cos\left(k_x + s_{\sigma}\,\phi_J\right) -2t\cos k_y - h s_{\sigma} - \mu,
\end{equation}
where $t_1=\sqrt{t^2+J^2}$ and $\phi_J = \tan^{-1}\!\left(\frac{J}{t}\right)$.

Microscopically, the model in Eq.~\eqref{eq:model} is an effective single-sublattice description of a parent lattice model, in which coplanar local moments rotating by $90^{\circ}$ at each nearest-neighbor step generate the $p$-wave exchange field experienced by itinerant electrons hopping between the two square sublattices~\cite{Kim2026a, Khodas2026}.

\subsection{Zero-momentum pairing states}

\begin{table}[t!]
\caption{Pairing channels are classified according to the $C_{2v}$ irreducible representations (irreps). The broken inversion symmetry merges singlet and triplet states into the same $A_1$ irrep. The exchange field breaks the fourfold rotation, so the bond singlets $\Delta_{s_x}$ and $\Delta_{s_y}$ --- which would otherwise combine into a single higher-dimensional irrep --- instead belong to $A_1$ individually. The table is adapted from Ref.~\cite{Kim2026a}.}
\label{tab:irreps}
\begin{ruledtabular}
\begin{tabular}{lllll}
Irrep & $\Delta_\alpha$ & $g_\alpha(\bm{k})$ & Parity & Source \\
\hline
A$_1$ & $\Delta_{s}$        & $g_s = 1$                 & even & $U$ \\
A$_1$ & $\Delta_{s_x}$      & $g_{s_x} = \cos k_x$      & even & $V$ \\
A$_1$ & $\Delta_{s_y}$      & $g_{s_y} = \cos k_y$      & even & $V$ \\
A$_1$ & $\Delta_{p_x}$      & $g_{p_x} = \sin k_x$      & odd  & $V$ \\
\hline
B$_2$ & $\Delta_{p_y}$      & $g_{p_y}=\sin k_y$        & odd  & $V$ \\
\end{tabular}
\end{ruledtabular}
\end{table}

To describe the superconducting states, we perform a self-consistent BdG analysis. The interaction term in Eq.~\eqref{eq:model} can be decomposed into pairing channels, each with its own form factor $g_{\alpha}(\bm{k})$. For the inter-spin zero-momentum pairing states---that is, pairing between up- and down-spin electrons---these channels are classified according to the irreducible representations (irreps) of the $C_{2v}$ point group, as summarized in Table~\ref{tab:irreps}. Two symmetry-allowed channels exist: the $A_1$ and $B_2$ states. The $A_1$ state is described by the mean-field BdG Hamiltonian
\begin{equation} \label{eq:H_BdG_A1}
\hat{H}_{\mathrm{BdG}}^{A_1} =\sum_{\bm{k}} \Psi^{\dagger}_{\bm{k}} \begin{pmatrix} \epsilon_{\uparrow}(\bm{k}) & \Delta_{A_1}(\bm{k})\\ [\Delta_{A_1}(\bm{k})]^{*} & -\epsilon_{\downarrow}(-\bm{k})\end{pmatrix} \Psi_{\bm{k}} + C,
\end{equation}
with the Nambu basis $\Psi_{\bm{k}}=\begin{pmatrix}c_{\bm{k}\uparrow},c_{-\bm{k}\downarrow}^{\dagger}\end{pmatrix}^{T}$. The constant $C$ collects the single-particle energy shift generated by normal ordering together with the terms quadratic in the pairing amplitudes, $C=\sum_{\bm{k}}\epsilon_{\dn}(\bm{k})+\frac{N}{U}|\Delta_{s}|^{2}+\frac{N}{2V}\sum_{\alpha=s_x,s_y,p_x}|\Delta_{\alpha}|^{2}$. The gap function $\Delta_{A_1}(\bm{k})$ comprises three even-parity components ($\alpha = s,\, s_x,\, s_y$) and one odd-parity component ($\alpha = p_x$),
\begin{equation} \label{eq:gap_fnc_A1}
    \Delta_{A_1}(\bm{k}) = \sum_{\alpha=s,s_x,s_y,p_x} \Delta_{\alpha}\, g_{\alpha}(\bm{k}).
\end{equation}
The exchange field breaks inversion symmetry and therefore mixes the even-parity spin-singlet and odd-parity spin-triplet channels belonging to the same irrep. Each pairing amplitude satisfies its own self-consistency condition,
\begin{equation}
\begin{aligned} \label{eq:gap_eq_A1}
\Delta_{s} = & \; -\frac{U}{N}\sum_{\bm{k}} \left \langle c_{-\bm{k}\dn}\, c_{\bm{k}\up} \right \rangle, \\
\Delta_{s_x} = & \; -\frac{2V}{N}\sum_{\bm{k}} \cos k_x\, \left \langle c_{-\bm{k}\dn}\, c_{\bm{k}\up} \right \rangle, \\
\Delta_{s_y} = & \; -\frac{2V}{N}\sum_{\bm{k}} \cos k_y\, \left \langle c_{-\bm{k}\dn}\, c_{\bm{k}\up} \right \rangle, \\
\Delta_{p_x} = & \; -\frac{2V}{N}\sum_{\bm{k}} \sin k_x\, \left \langle c_{-\bm{k}\dn}\, c_{\bm{k}\up} \right \rangle,
\end{aligned}
\end{equation}
where $N$ is the number of sites.

The $B_2$ state is described by the BdG Hamiltonian
\begin{equation} \label{eq:H_BdG_B2}
\hat{H}_{\mathrm{BdG}}^{B_2} =\sum_{\bm{k}} \Psi^{\dagger}_{\bm{k}}
\begin{pmatrix}
\epsilon_{\uparrow}(\bm{k}) & \Delta_{B_2}(\bm{k})\\
[\Delta_{B_2}(\bm{k})]^{*} & -\epsilon_{\downarrow}(-\bm{k})
\end{pmatrix}
\Psi_{\bm{k}} + C,
\end{equation}
with the same Nambu spinor as in the $A_1$ case and the constant $C=\sum_{\bm{k}}\epsilon_{\dn}(\bm{k})+\frac{N}{2V}|\Delta_{p_y}|^{2}$. The gap function reads
\begin{equation} \label{eq:gap_fnc_B2}
\Delta_{B_2}(\bm{k}) = \Delta_{p_y}g_{p_y}(\bm{k}),
\end{equation}
where $\Delta_{p_y}$ is determined self-consistently as
\begin{equation} \label{eq:gap_eq_B2}
\Delta_{p_y} = -\frac{2V}{N}\sum_{\bm{k}} \sin k_y\, \left \langle c_{-\bm{k}\dn}\, c_{\bm{k}\up} \right \rangle.
\end{equation}

\subsection{Finite-momentum pairing states}

We assume an equal-spin configuration for the FF states, which restricts the pairing to the odd-parity components $\alpha\in\{p_x,p_y\}$. We choose the pair momentum as
\begin{equation} \label{eq:pair_momentum}
\bm{Q}_\sigma = -2s_{\sigma}\phi_{J}\,\hat{\bm k}_x \equiv Q_\sigma\hat{\bm k}_x.
\end{equation}
Equal-spin pairing at a finite pair momentum $\bm Q_\sigma\parallel\hat{\bm k}_x$ reduces the relevant symmetry group from $C_{2v}$ to $C_s=\{E,M_y\}$, where $M_y$ denotes the mirror operation $k_y\rightarrow-k_y$ that leaves $\bm{Q}_\sigma$ invariant. Under this group, the form factors $g_{p_x}(\bm{k})=\sin k_x$ and $g_{p_y}(\bm{k})=\sin k_y$ belong to the irreps $A'$ and $A''$, respectively, so the two associated states, hereafter referred to as the FF-$p_x$ and FF-$p_y$ states, do not mix and may be treated separately.
They are described by the BdG Hamiltonian
\begin{equation} \label{eq:H_BdG_FF}
\hat{H}_{\mathrm{BdG}}^{\mathrm{FF}\text{-}\alpha} = \frac{1}{2} \sum_{\bm{k},\sigma} \Psi^{\dagger}_{\bm{k},\sigma} \begin{pmatrix} \tilde{\epsilon}_{\sigma}(\bm{k}) & \Delta_{\alpha;\sigma}^{\mathrm{FF}}(\bm{k})\\ \left[\Delta_{\alpha;\sigma}^{\mathrm{FF}}(\bm{k})\right]^{*} & -\tilde{\epsilon}_{\sigma}(-\bm{k}) \end{pmatrix} \Psi_{\bm{k},\sigma} + C,
\end{equation}
where $\alpha=p_x$ ($p_y$) corresponds to the FF-$p_x$ (FF-$p_y$) state. Here, the Nambu basis is $\Psi_{\bm{k},\sigma}=\bigl(c_{\bm{k}+\frac{\bm{Q}_\sigma}{2},\sigma},\,c^{\dagger}_{-\bm{k}+\frac{\bm{Q}_\sigma}{2},\sigma}\bigr)^{\mathrm{T}}$ and the constant is $C=\frac12\sum_{\bm{k},\sigma}\epsilon_{\sigma}(\bm{k})+\frac{N}{4V}\sum_{\sigma}\bigl|\Delta_{\alpha;\sigma}^{\mathrm{FF}}\bigr|^2$. The gap function is
\begin{equation}
    \Delta_{\alpha;\sigma}^{\mathrm{FF}}(\bm{k}) =\Delta_{\alpha;\sigma}^{\mathrm{FF}}\,g_{\alpha}(\bm{k}),
\end{equation}
and the shifted dispersion $\tilde{\epsilon}_\sigma(\bm{k}) \equiv \epsilon_{\sigma}\big(\bm{k}+\frac{\bm{Q}_\sigma}{2}\big)$ reads
\begin{equation} \label{eq:epsilon_tilde}
    \tilde{\epsilon}_\sigma(\bm{k}) = -2t_1 \cos k_x  -2t\cos k_y - hs_{\sigma} - \mu.
\end{equation}
The pairing amplitude is determined self-consistently by
\begin{equation} \label{eq:gap_eq_FF}
\Delta_{\alpha;\sigma}^{\mathrm{FF}} = -\frac{2V}{N}\sum_{\bm{k}} g_{\alpha}(\bm{k}) \left\langle c_{-\bm{k}+\frac{\bm{Q}_{\sigma}}{2},\sigma}\, c_{\bm{k}+\frac{\bm{Q}_\sigma}{2},\sigma}\right\rangle.
\end{equation}

We emphasize that the choice of pair momentum in Eq.~\eqref{eq:pair_momentum} is essential for the energetic stabilization of the FF states. With this choice, the two electrons forming a Cooper pair have exactly the same energy [see Fig.~\ref{fig1}(a)],
\begin{equation} \label{eq:equal_energy}
\tilde{\epsilon}_{\sigma}(\bm{k}) = \tilde{\epsilon}_{\sigma}(-\bm{k}).
\end{equation}
In general, the denominator of the gap equation [Eq.~\eqref{eq:gap_eq_FF_explicit}] involves the symmetrized dispersion $\tilde{\epsilon}_\sigma^S(\bm{k})=\frac{1}{2}[\tilde{\epsilon}_{\sigma}(\bm{k})+\tilde{\epsilon}_{\sigma}(-\bm{k})]$ through $\sqrt{\tilde{\epsilon}_\sigma^S(\bm{k})^2+\bigl|\Delta_{\alpha;\sigma}^{\mathrm{FF}} g_{\alpha}(\bm{k})\bigr|^2}$. Under the equal-energy condition of Eq.~\eqref{eq:equal_energy}, this reduces to $\sqrt{\tilde{\epsilon}_\sigma(\bm{k})^2+\bigl|\Delta_{\alpha;\sigma}^{\mathrm{FF}} g_{\alpha}(\bm{k})\bigr|^2}$, governed by the single dispersion $\tilde{\epsilon}_\sigma(\bm{k})$. The zero-energy contour of this dispersion therefore coincides with the entire shifted FS, supporting a substantial pairing amplitude in this channel. By contrast, for a general $\bm{Q}_\sigma$ the two paired electrons have different energies $\tilde{\epsilon}_{\sigma}(\bm{k})$ and $\tilde{\epsilon}_{\sigma}(-\bm{k})$, and the zero-energy contour of $\tilde{\epsilon}_\sigma^S$ no longer covers a full FS. The pairing amplitude is therefore suppressed whenever $\bm{Q}_\sigma$ deviates from Eq.~\eqref{eq:pair_momentum}, indicating that this choice is the optimal pair momentum for stabilizing the FF states.

In total, we investigate four superconducting states: the equal-spin FF-$p_x$ and FF-$p_y$ states [Fig.~\ref{fig1}(a)], and the two inter-spin $A_1$ and $B_2$ states with zero pair momentum [Fig.~\ref{fig1}(b)]. These four channels cover the pairing configurations in which every electron on the FSs can participate, thereby potentially maximizing the pairing amplitude.

\section{Emergence of FF states}

\subsection{Zero-field phase diagrams} \label{sec:zero_field_phase_diagram}

\begin{figure*}[t!]
    \centering
    \includegraphics[width=.8\textwidth]{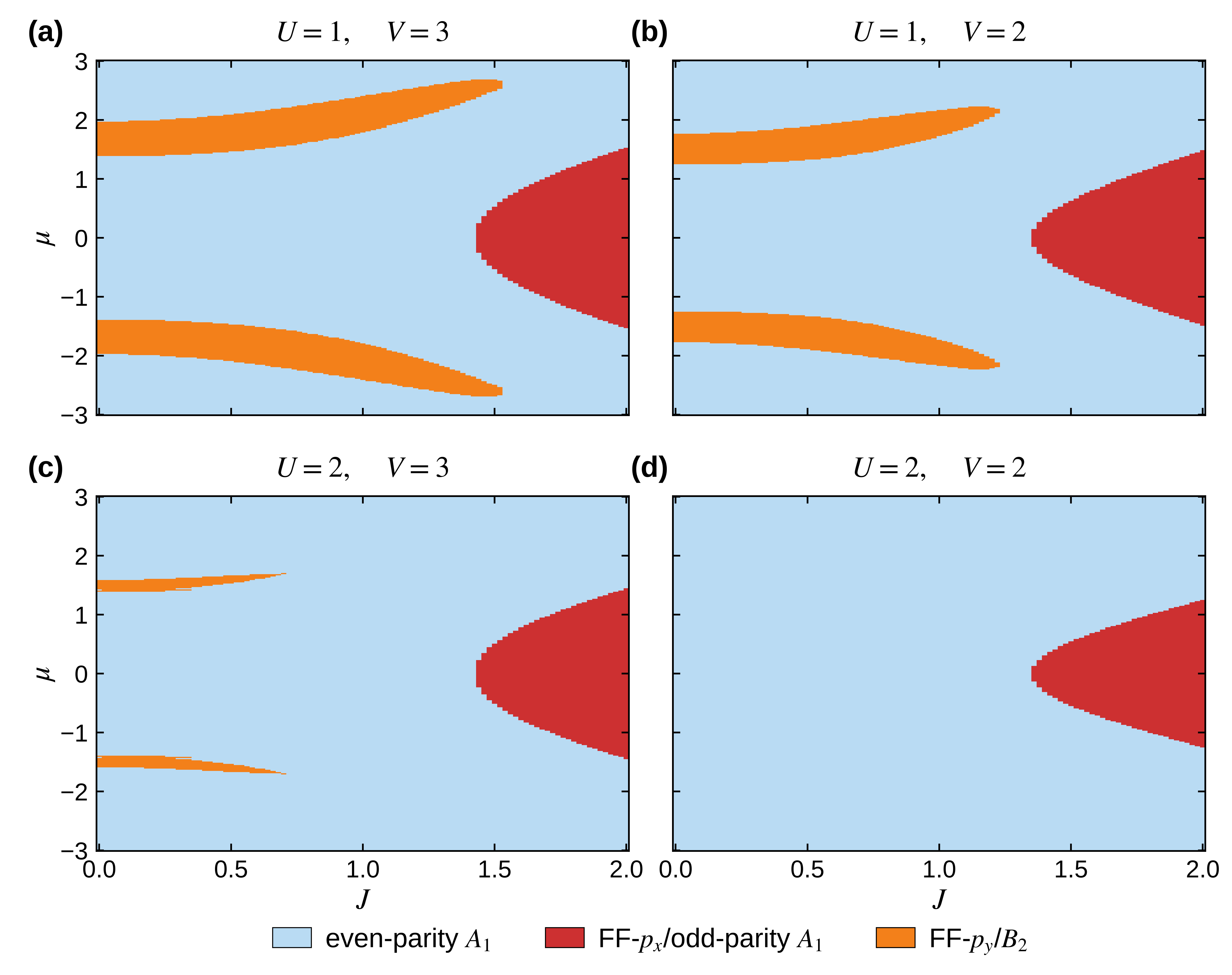} \caption{\textbf{Emergence of FF states.} (a)--(d) Phase diagrams of the pairing ground state as functions of the $p$-wave exchange field strength $J$ and the chemical potential $\mu$. Light blue denotes the even-parity $A_1$ phase, red the degenerate FF-$p_x$/odd-parity $A_1$ phase, and orange the degenerate FF-$p_y$/$B_2$ phase. The interaction strengths are (a) $(U,V)=(1,3)$, (b) $(1,2)$, (c) $(2,3)$, and (d) $(2,2)$.}
    \label{fig2}
\end{figure*}

We determine the energetically favored superconducting state among the four candidate pairing states---$A_1$, $B_2$, FF-$p_x$, and FF-$p_y$---over a range of parameters ($\mu$, $J$, $U$, and $V$) by numerically solving the self-consistent gap equations at zero temperature. The resulting pairing amplitudes are substituted into the corresponding ground-state energy expressions, and the state with the lowest energy is identified as the superconducting ground state, from which the phase diagrams are constructed. The explicit gap equations, ground-state energy expressions, and the numerical procedure are provided in the Methods section.

Figure~\ref{fig2} displays phase diagrams of the superconducting ground state in the $J$--$\mu$ plane for four combinations of $(U,V)$. At $(U,V)=(1,3)$, the $A_1$ phase occupies most of the parameter space [Fig.~\ref{fig2}(a)]. The FF-$p_x$ phase occupies a region centered at $\mu=0$, which sets in at $J\simeq1.44$ and widens with increasing $J$, reaching $|\mu|\simeq1.5$ at $J=2$; this region is symmetric under $\mu\leftrightarrow-\mu$, as required by the particle-hole symmetry of the bipartite square lattice with nearest-neighbor hopping. By contrast, the FF-$p_y$ phase occupies narrow regions at intermediate chemical potential, $1.4\lesssim|\mu|\lesssim2.0$ at $J=0$, which drift toward larger $|\mu|$ as $J$ increases and terminate at $J\simeq1.5$. Both the FF-$p_x$ and FF-$p_y$ phases are degenerate with their zero-momentum counterparts, namely the $A_1$ and $B_2$ phases, respectively.

Qualitatively similar phase diagrams are obtained for other $(U,V)$ combinations. Reducing $V$ from $3$ to $2$ at the same $U$ shrinks the FF-$p_y$/$B_2$ regions, while the FF-$p_x$/$A_1$ region keeps its onset at $J\simeq1.4$ and only its $\mu$ range becomes slightly narrower [Fig.~\ref{fig2}(b)]. Raising $U$ from $1$ to $2$ at the same $V$ suppresses the former far more strongly, whereas the latter again keeps its $J$ onset and its $\mu$ range narrows further [Fig.~\ref{fig2}(c)]. When both changes are combined at $(U,V)=(2,2)$, the FF-$p_y$/$B_2$ phase disappears altogether, whereas the FF-$p_x$/$A_1$ phase survives with a similar onset at $J\simeq1.4$ and a narrower $\mu$ range [Fig.~\ref{fig2}(d)]. These results indicate that the degenerate FF-$p_x$/$A_1$ phase persists at large $J$ near $\mu=0$ largely independently of $U$ and $V$ over the range examined, whereas its FF-$p_y$/$B_2$ counterpart requires a sufficiently large $V$ and a sufficiently small $U$ and is suppressed otherwise. Increasing $U$ further to $3$ eliminates either the FF-$p_y$ phase alone or both the FF-$p_y$ and FF-$p_x$ phases, depending on the value of $V$. A detailed analysis is given in Supplementary Note~1.

\begin{figure*}
    \centering
    \includegraphics[width=.8\textwidth]{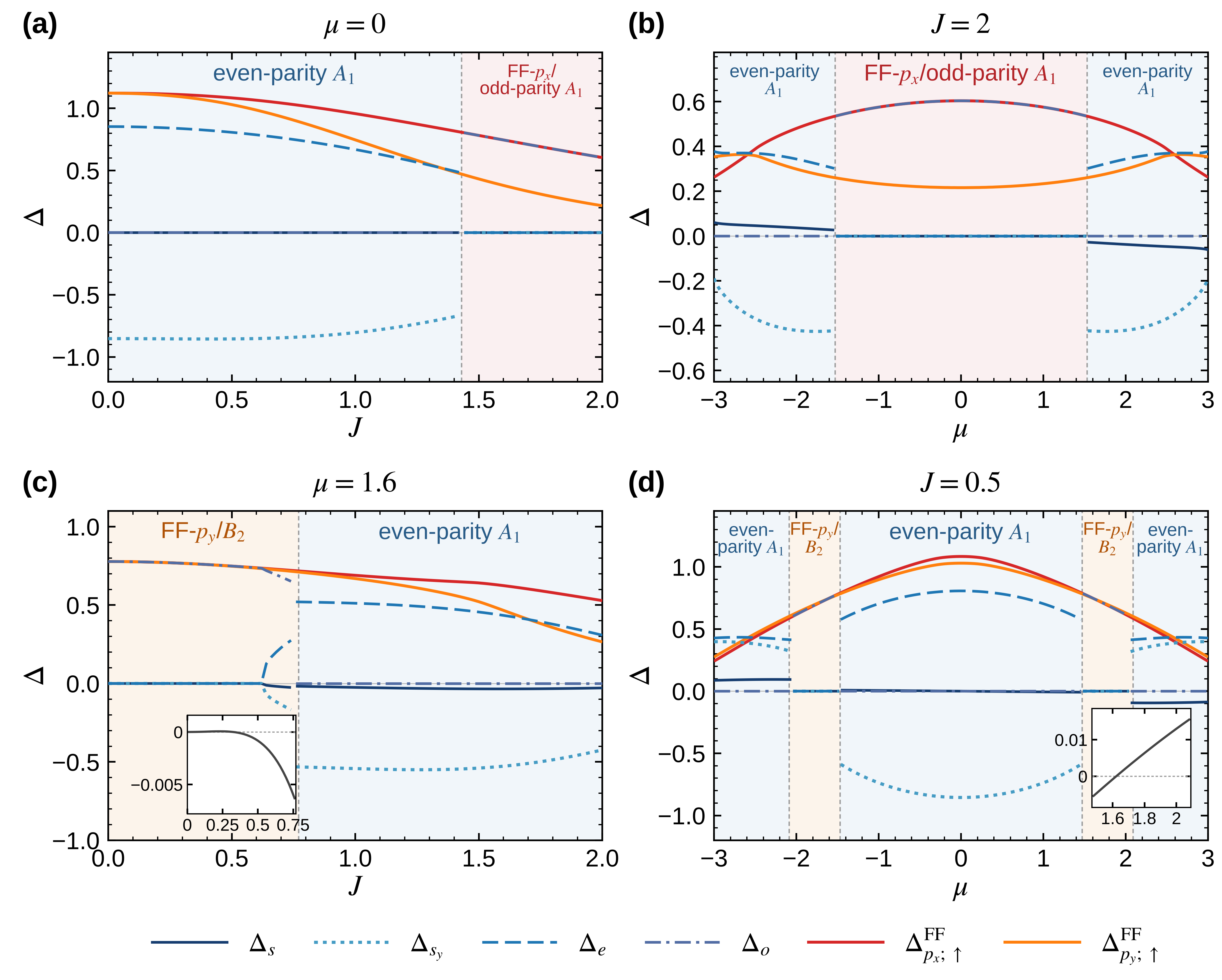}
    \caption{\textbf{Distribution of pairing amplitude components.} (a), (c) Pairing amplitudes as a function of $J$ for (a) $\mu=0$ and (c) $\mu=1.6$. (b), (d) Pairing amplitudes as a function of $\mu$ for (b) $J=2$ and (d) $J=0.5$. The parameters $(U,V)=(1,3)$ are used throughout, as in Fig.~\ref{fig2}(a). The curves denote the $s$-wave ($\Delta_{s}$, blue solid), $s_y$-wave ($\Delta_{s_y}$, blue dotted), even-parity ($\Delta_{e}$, blue dashed), and odd-parity ($\Delta_{o}$, blue dot-dashed) components of the $A_1$ state, together with the FF amplitudes $\Delta^{\mathrm{FF}}_{p_x;\up}$ (red) and $\Delta^{\mathrm{FF}}_{p_y;\up}$ (orange). In (c) and (d), insets display the difference $\Delta^{\mathrm{FF}}_{p_y;\up}-\Delta^{\mathrm{FF}}_{p_x;\up}$ within the FF-$p_y$/$B_2$ phase. In each panel, the stabilized ground-state phase is annotated within its region, with the regions separated by dashed lines.}
    \label{fig3}
\end{figure*}

We find that within the $A_1$ phase, the pairing components $\Delta_{s_x}$ and $\Delta_{p_x}$ of the $A_1$ gap function are locked into a vector form $(\Delta_{s_x},\Delta_{p_x})=\Delta_e (\cos{\phi_J}, -\sin{\phi_J})\equiv\bm{\Delta}_{x}^\mathrm{even}$, whose orientation is determined by the momentum shift $\phi_J$. This locking combines the two form factors $g_{s_x}(\bm{k})$ and $g_{p_x}(\bm{k})$ into the single form factor $\cos{(k_x+\phi_J)}$, which carries the same momentum shift as the energy dispersion and is even under the reflection $k_x \rightarrow -k_x-2\phi_J$, giving the even-parity gap function
\begin{equation}
    \Delta_{A_1}^\mathrm{even}(\bm{k}) = \Delta_{s} + \Delta_{s_y}\cos{k_y} + \Delta_e \cos{(k_x+\phi_J)}.\label{eq:A1_even_branch}
\end{equation} 
In the degenerate FF-$p_x$/$A_1$ phase, by contrast, $\Delta_{s_x}$ and $\Delta_{p_x}$ are locked into the vector form $(\Delta_{s_x},\Delta_{p_x})=\Delta_o (\sin{\phi_J}, \cos{\phi_J})\equiv\bm{\Delta}_{x}^\mathrm{odd}$, while $\Delta_s=\Delta_{s_y}=0$. The two form factors then combine into $\sin{(k_x+\phi_J)}$, which is odd under the same reflection, leading to the odd-parity gap function
\begin{equation}
    \Delta_{A_1}^\mathrm{odd}(\bm{k}) = \Delta_o \sin{(k_x+\phi_J)}.\label{eq:A1_odd_branch}
\end{equation} 
The pairing vectors $\bm{\Delta}_{x}^\mathrm{even}$ and $\bm{\Delta}_{x}^\mathrm{odd}$ are eigenvectors of the reflection matrix
\begin{equation}
R_J = \begin{pmatrix} \cos{2\phi_J} & -\sin{2\phi_J} \\ -\sin{2\phi_J} & -\cos{2\phi_J} \end{pmatrix},
\label{eq:A1_reflection}
\end{equation}
with eigenvalues $+1$ and $-1$, respectively, which define their parities. Each is therefore mapped onto itself under $R_J$, up to an overall sign for the odd-parity vector that can be removed by a global phase rotation, and thus has no symmetry-related partner. This is in contrast to a general mixed form, $\bm{\Delta}_x^\mathrm{mixed}=\bm{\Delta}_x^\mathrm{even}+\bm{\Delta}_x^\mathrm{odd}$, with the gap function $\Delta_{A_1}^\mathrm{mixed}(\bm{k}) = \Delta_{s} + \Delta_{s_y}\cos{k_y} + \Delta_e \cos{(k_x+\phi_J)}+\Delta_o \sin{(k_x+\phi_J)}$, which is not invariant under $R_J$ and is always accompanied by a degenerate partner obtained by applying $R_J$ to it. A detailed analysis of this degeneracy is provided in Supplementary Note~2.

Figure~\ref{fig3} shows the four $A_1$ amplitudes $\Delta_s$, $\Delta_{s_y}$, $\Delta_e$, and $\Delta_o$, together with the two FF amplitudes, along $J$-cuts at fixed $\mu$ and $\mu$-cuts at fixed $J$, for $(U,V)=(1,3)$ as in Fig.~\ref{fig2}(a). The $B_2$ amplitude is omitted because it always equals that of the FF-$p_y$ state (see Supplementary Note~2), and only the spin-up components of the FF states are shown because the spin-down components are identical. Along the $J$-cut at $\mu=0$, $\Delta_e$ and $\Delta_{s_y}$ are nonzero within the $A_1$ phase whereas $\Delta_o$ remains zero, consistent with the even-parity $A_1$ state [Fig.~\ref{fig3}(a)]. In the degenerate FF-$p_x$/$A_1$ phase, by contrast, $\Delta_o$ is the only nonzero component, and its magnitude is identical to that of $\Delta_{p_x;\uparrow}^{\mathrm{FF}}$. This is not a coincidence but reflects the degeneracy of the two states within this phase, which will be elucidated in Sec.~\ref{sec:selection_mechanism}. Similar patterns are observed along the $\mu$-cut at $J=2$ [Fig.~\ref{fig3}(b)]. Within the degenerate FF-$p_x$/$A_1$ phase, $\Delta_o$ is the only nonzero component and equals $\Delta_{p_x;\uparrow}^{\mathrm{FF}}$, whereas within the $A_1$ phase, $\Delta_o$ vanishes and $\Delta_s$, $\Delta_{s_y}$, and $\Delta_e$ are nonzero.

The $J$-cut at $\mu=1.6$ shows that, within the degenerate FF-$p_y$/$B_2$ phase, $\Delta_{p_x;\uparrow}^{\mathrm{FF}}$ exceeds $\Delta_{p_y;\uparrow}^{\mathrm{FF}}$ by a slight margin [Fig.~\ref{fig3}(c), inset]. Within the same phase, the $A_1$ gap function is of the odd-parity type $\bm{\Delta}_x^\mathrm{odd}$ for $J<0.63$, whereas it takes the mixed form $\bm{\Delta}_x^\mathrm{mixed}$ for $0.63\leq J<0.76$. For $J\geq0.78$, where the $A_1$ phase is stabilized, the $A_1$ gap function is of the even-parity type $\bm{\Delta}_x^\mathrm{even}$. These observations indicate that the odd-parity and mixed forms appear only within the degenerate phases, whereas the even-parity form characterizes the $A_1$ phase. Similar behavior is observed along the $\mu$-cut at $J=0.5$ [Fig.~\ref{fig3}(d)]. Within the degenerate FF-$p_y$/$B_2$ windows at $1.46<|\mu|<2.10$, $\Delta_{p_y;\uparrow}^{\mathrm{FF}}$ exceeds $\Delta_{p_x;\uparrow}^{\mathrm{FF}}$ by a small margin [Fig.~\ref{fig3}(d), inset], and the $A_1$ gap function is of the odd-parity type, whereas it is of the even-parity type outside these windows. It is worth mentioning that $\Delta_s$ remains the smallest component of the even-parity $A_1$ gap function over most of the $J$--$\mu$ plane, which we attribute to the small values of $U$ ($U=1,2$) considered here, whereas for $U=3$ it can become comparable to the other components, as detailed in Supplementary Note~1.

\subsection{Pairing-channel selection mechanism} \label{sec:selection_mechanism}

\begin{figure*}[t!]
    \centering
    \includegraphics[width=\linewidth]{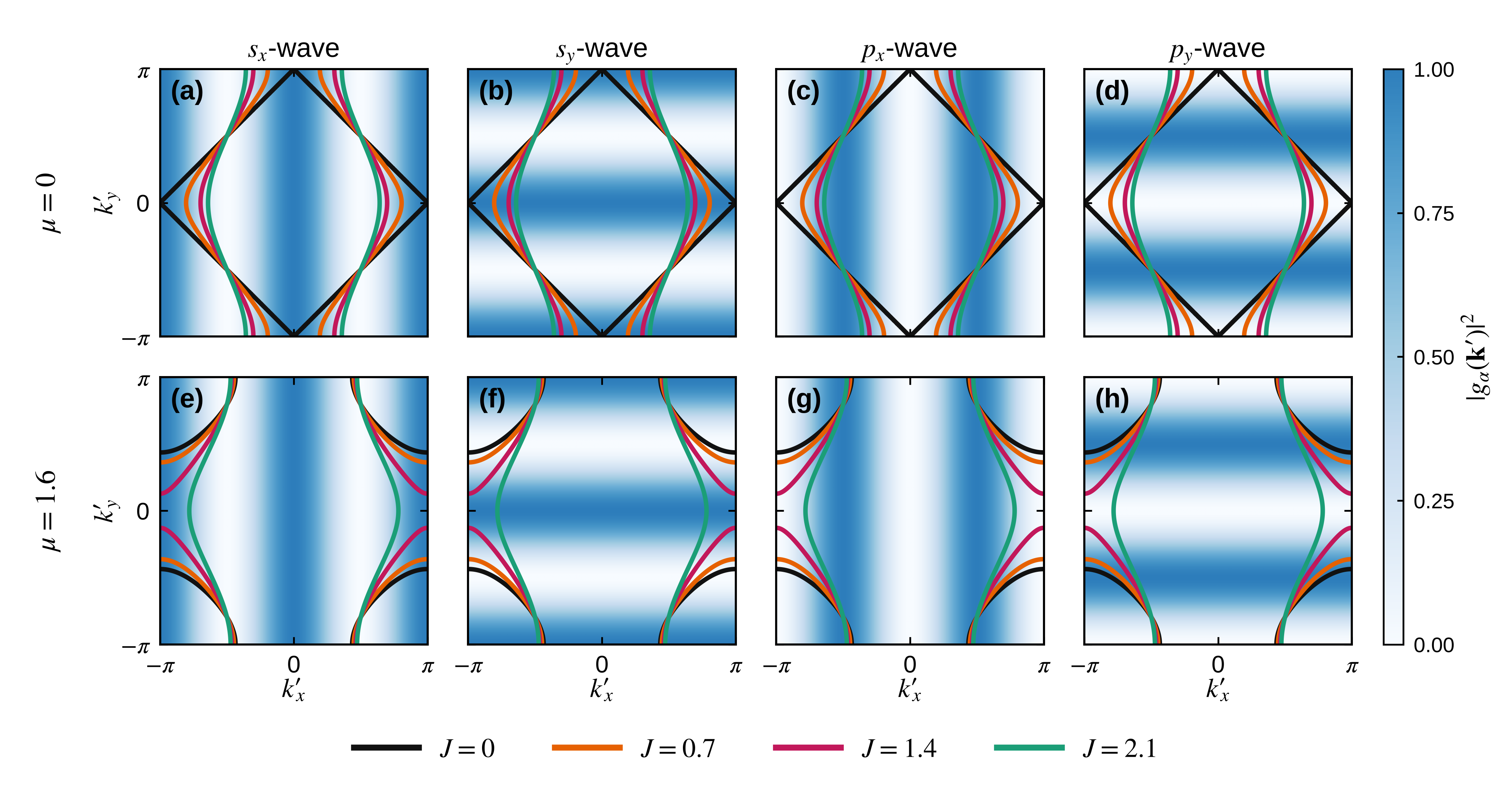}
    \caption{\textbf{Selection of pairing channels.} Squared form factors $|g_\alpha(\bm{k}')|^2$ (color scale) overlaid with the zero-energy contour $\tilde{\epsilon}(\bm{k}')=0$ (lines) at the exchange field strengths $J=0$, $0.7$, $1.4$, and $2.1$ (black, orange, magenta, and green, respectively). (a)--(d) $\mu=0$, for the (a) $s_x$-, (b) $s_y$-, (c) $p_x$-, and (d) $p_y$-wave form factors. (e)--(h) Same as (a)--(d), but at $\mu=1.6$.}
    \label{fig4}
\end{figure*}

Having established the phase diagrams of the FF states and their degenerate zero pair-momentum counterparts, we now examine the mechanism that selects the pairing channel in each region of the $J$--$\mu$ plane. To this end, we consider reduced gap equations for the even-parity $A_1$, odd-parity $A_1$, and $B_2$ states at $h=0$,
\begin{align}
    \Delta_e = & \; \frac{V}{N} \sum_{\bm{k}'}\frac{\bigl[\Delta_{s_y}g_{s_y}(\bm{k}') + \Delta_e g_{s_x}(\bm{k}')\bigr]g_{s_x}(\bm{k}')}{\sqrt{\tilde{\epsilon}(\bm{k}')^2+ |\Delta_{s_y}g_{s_y}(\bm{k}') + \Delta_e g_{s_x}(\bm{k}')|^2}}, \label{eq:gap_eq_A1_even_e_explicit_h0} \\
    \Delta_{s_y} = & \; \frac{V}{N} \sum_{\bm{k}'}\frac{\bigl[\Delta_{s_y}g_{s_y}(\bm{k}') + \Delta_e g_{s_x}(\bm{k}')\bigr]g_{s_y}(\bm{k}')}{\sqrt{\tilde{\epsilon}(\bm{k}')^2+ |\Delta_{s_y}g_{s_y}(\bm{k}') + \Delta_e g_{s_x}(\bm{k}')|^2}}, \label{eq:gap_eq_A1_even_s_y_explicit_h0} \\
    \Delta_o = & \; \frac{V}{N} \sum_{\bm{k}'}\frac{\Delta_o\, g_{p_x}(\bm{k}')^2}{\sqrt{\tilde{\epsilon}(\bm{k}')^2+ |\Delta_o\, g_{p_x}(\bm{k}')|^2}}, \label{eq:gap_eq_A1_odd_explicit_h0} \\
    \Delta_{p_y} = & \; \frac{V}{N} \sum_{\bm{k}'}\frac{\Delta_{p_y}\, g_{p_y}(\bm{k}')^2}{\sqrt{\tilde{\epsilon}(\bm{k}')^2+ |\Delta_{p_y}\, g_{p_y}(\bm{k}')|^2}},\label{eq:gap_eq_B2_explicit_h0}
\end{align}
where the dispersion is
\begin{equation}
    \tilde{\epsilon}(\bm{k}')=-2t_1\cos{k_x'}-2t\cos{k_y'}-\mu,
\end{equation}
the form factors are $g_{s_x}=\cos k_x'$, $g_{s_y}=\cos k_y'$, $g_{p_x}=\sin k_x'$, and $g_{p_y}=\sin k_y'$, and the momentum $\bm{k}'$ represents the shifted momentum coordinates
\begin{equation} \label{eq:shifted_momentum}
k_x' = k_x + \phi_J, \qquad k_y' = k_y.
\end{equation}
The derivation of the reduced gap equations is given in Supplementary Note~3. Equations~\eqref{eq:gap_eq_A1_even_e_explicit_h0} and \eqref{eq:gap_eq_A1_even_s_y_explicit_h0} govern the even-parity $A_1$ pairing amplitudes, where we have neglected $\Delta_s$ for simplicity. A fully self-consistent treatment would also require the gap equation for $\Delta_s$, but the omission is supported by the small magnitude of $\Delta_s$ in the self-consistent solutions (Fig.~\ref{fig3}). Equations~\eqref{eq:gap_eq_A1_odd_explicit_h0} and \eqref{eq:gap_eq_B2_explicit_h0} govern the odd-parity $A_1$ and $B_2$ amplitudes, respectively. The momentum summation is dominated by the vicinity of the contour $\tilde{\epsilon}(\bm{k}')=0$, where the summand is weighted by the corresponding form factor. The pairing channel selected for given parameters is therefore governed by the overlap between the contour and the form factors. 

To visualize this overlap, we plot the zero-energy contour $\tilde{\epsilon}(\bm{k}')=0$ together with the momentum-space distributions of the form factors $g_{s_x}(\bm{k}')$, $g_{s_y}(\bm{k}')$, $g_{p_x}(\bm{k}')$, and $g_{p_y}(\bm{k}')$, as illustrated in Fig.~\ref{fig4}. At $\mu=0$ and $J=0$, the contour passes through the van Hove points $(k_x',k_y')=(\pi,0)$ and $(0,\pi)$, where the density of states diverges. At these points, $g_{s_x}(\bm{k}')$ and $g_{s_y}(\bm{k}')$ reach their antinodes while $g_{p_x}(\bm{k}')$ and $g_{p_y}(\bm{k}')$ vanish. In addition, the fourfold rotation symmetry ensures that $g_{s_x}(\bm{k}')$ and $g_{s_y}(\bm{k}')$ have identical overlaps with the contour, and likewise for $g_{p_x}(\bm{k}')$ and $g_{p_y}(\bm{k}')$ [Figs.~\ref{fig4}(a)--(d)]. The overlaps entering the gap equations \eqref{eq:gap_eq_A1_even_e_explicit_h0}--\eqref{eq:gap_eq_B2_explicit_h0} are therefore larger for $g_{s_x}(\bm{k}')$ and $g_{s_y}(\bm{k}')$ than for $g_{p_x}(\bm{k}')$ and $g_{p_y}(\bm{k}')$. This accounts for the stabilization of the even-parity $A_1$ state, built from the cosine channels, at small $J$ [Fig.~\ref{fig3}(a)]. In this regime, the two cosine channels have comparable overlaps, and the even-parity gap takes the $d$-wave-like form $\Delta_e\cos k_x' +\Delta_{s_y}\cos k_y'$ with $\Delta_{s_y}\approx-\Delta_e$, which maximizes the gap magnitude at the two van Hove points, $|\Delta_{A_1}^\mathrm{even}|\approx2|\Delta_e|$. At larger $J$, the contour moves away from the van Hove points and the fourfold rotation symmetry is broken, so that the four overlaps all become different. Specifically, at $J=0.7$, the contour has greater overlap with $g_{s_y}(\bm{k}')$ than with $g_{s_x}(\bm{k}')$. This accounts for the observed growth of the $s_y$ component in the even-parity $A_1$ state [Fig.~\ref{fig3}(a)]. As $J$ increases further, the exchange field elongates the contour along $k_y'$ into nearly straight segments near $k_x'=\pm\pi/2$, which are the nodes of $g_{s_x}(\bm{k}')$ and the antinodes of $g_{p_x}(\bm{k}')$ [Figs.~\ref{fig4}(a) and (c)]. Consequently, at $J\geq1.44$, the odd-parity $A_1$ state, governed by $g_{p_x}(\bm{k}')$, becomes energetically favored over the even-parity one.

At $\mu=1.6$ and $J=0$, the contour runs through the antinodes of the sine form factors $g_{p_x}(\bm{k}')$ and $g_{p_y}(\bm{k}')$ [Figs.~\ref{fig4}(g) and (h)] and close to the nodes of the cosine form factors $g_{s_x}(\bm{k}')$ and $g_{s_y}(\bm{k}')$ [Figs.~\ref{fig4}(e) and (f)]. The sine channels therefore carry the larger overlap, so that the odd-parity $A_1$ and $B_2$ states, governed by the sine channels, are favored over the even-parity $A_1$ state, governed by the cosine channels. Introducing a small $J$ enhances the overlap with the antinode of $g_{p_y}(\bm{k}')$ while reducing that of $g_{p_x}(\bm{k}')$ [Figs.~\ref{fig4}(g) and (h)], stabilizing the $B_2$ state for $J<0.78$ [Fig.~\ref{fig3}(c)]. In the same $J$ interval, the overlap with $g_{p_x}(\bm{k}')$ still exceeds those with $g_{s_x}(\bm{k}')$ and $g_{s_y}(\bm{k}')$ [Figs.~\ref{fig4}(e)--(g)], so that the odd-parity form is favored over the even-parity one within the $A_1$ state [Fig.~\ref{fig3}(c)]. Finally, at $J=1.4$ and $2.1$, the dominance of $g_{p_y}(\bm{k}')$ is lost and the contour approaches the van Hove points, where both $g_{s_x}(\bm{k}')$ and $g_{s_y}(\bm{k}')$ reach their antinodes [Figs.~\ref{fig4}(e) and (f)], so that the even-parity $A_1$ state becomes the energetically favored one [Fig.~\ref{fig3}(c)].

We note that the odd-parity $A_1$ gap equation~\eqref{eq:gap_eq_A1_odd_explicit_h0} coincides with the FF-$p_x$ gap equation~\eqref{eq:gap_eq_FF_explicit} with $\alpha=p_x$. The coincidence arises because the global shift $k_x \rightarrow k_x-\phi_J$ removes $\phi_J$ from the $A_1$ gap equation and leaves it in the same form as the FF one, in which the pair momentum plays the same role. This equivalence is understood in terms of the position- and spin-dependent gauge transformation,
\begin{equation} \label{eq:gauge}
d_{i\sigma} = e^{\,i\,s_{\sigma}\phi_J x_i}\,c_{i\sigma}, ~ d_{\bm{k}'\sigma} = c_{\bm{k}\sigma}, ~ \bm{k}'=\bm{k}+s_\sigma\phi_J\hat{x}.
\end{equation}
Under this transformation, the $A_1$ BdG Hamiltonian~\eqref{eq:H_BdG_A1} with the odd-parity gap function~\eqref{eq:A1_odd_branch} becomes
\begin{equation}
\hat{H}^{A_1,\mathrm{odd}}_{\mathrm{BdG}}=\sum_{\bm{k}'}\Phi^{\dagger}_{\bm{k}'} \begin{pmatrix}\tilde{\epsilon}(\bm{k}') & \Delta_o\sin{k_x'}\\ \Delta_o^*\sin{k_x'} & -\tilde{\epsilon}(\bm{k}')
\end{pmatrix} \Phi_{\bm{k}'}+C,
\label{eq:BdG_A1_gauge}
\end{equation}
where $\Phi_{\bm{k}'} = (d_{\bm{k}'\uparrow},\, d^{\dagger}_{-\bm{k}'\downarrow})^T$. Equation~\eqref{eq:BdG_A1_gauge} is identical to the FF-$p_x$ BdG Hamiltonian~\eqref{eq:H_BdG_FF} apart from the spin structure of the Nambu spinor. This difference in the spin structure, however, is immaterial at $h=0$, where the two spin sectors of the FF-$p_x$ BdG Hamiltonian are equivalent, and the two problems therefore reduce to one and the same eigenvalue problem. By the same token, the $B_2$ gap equation~\eqref{eq:gap_eq_B2_explicit_h0} coincides with the FF-$p_y$ gap equation~\eqref{eq:gap_eq_FF_explicit}. The derivation given here further implies that the ground-state energies of the odd-parity $A_1$ and $B_2$ states agree with those of the FF-$p_x$ and FF-$p_y$ states, and a rigorous proof is provided in Supplementary Note~2. The correspondence established here indicates that the odd-parity $A_1$ and $B_2$ states are exactly degenerate with the FF-$p_x$ and FF-$p_y$ states, as reflected in Fig.~\ref{fig2}.

Taken together, the overlaps between the energy contour and the form factors examined above indicate that the alignment of the energy contour with the antinodes of $g_{p_x}(\bm{k}')$ at large $J$ and $\mu=0$ is the main driver for the stabilization of the odd-parity $A_1$ state in this parameter regime. By contrast, the overlap with $g_{p_y}(\bm{k}')$ is largest at small $J$ and $\mu \approx 1.6$, favoring the $B_2$ state instead. Because the FF-$p_x$ and FF-$p_y$ states are degenerate with the odd-parity $A_1$ and $B_2$ states at $h=0$, the same overlaps account for the stabilization of the FF-$p_x$ and FF-$p_y$ states in their respective regions.

While the discussion in this section has assumed $U=0$, the mechanism remains valid in the presence of $U>0$, where a finite $U$ directly enhances the $s$-wave component $\Delta_s$ of the even-parity $A_1$ state. As a result, the even-parity $A_1$ phase region widens at the expense of the FF-$p_x$ phase in the phase diagram [compare, e.g., Figs.~\ref{fig2}(a) and \ref{fig2}(c)]. This microscopic picture is also consistent with the observation that the $J$ threshold of the FF-$p_x$/even-parity $A_1$ phase boundary is nearly independent of $V$ [Figs.~\ref{fig2}(a) and \ref{fig2}(b)], which suggests that the threshold is set by the overlap between the contour and the form factors rather than by the strength of $V$. In contrast, the decrease of the $\mu$ threshold with $U$ can be attributed to the growth of $\Delta_s$ driven by $U$, which lies outside the mechanism discussed above.

\subsection{Stabilization of FF states via Zeeman effect} \label{sec:stabilization_FF_states}

\begin{figure*}[t!]
    \centering
    \includegraphics[width=.8\textwidth]{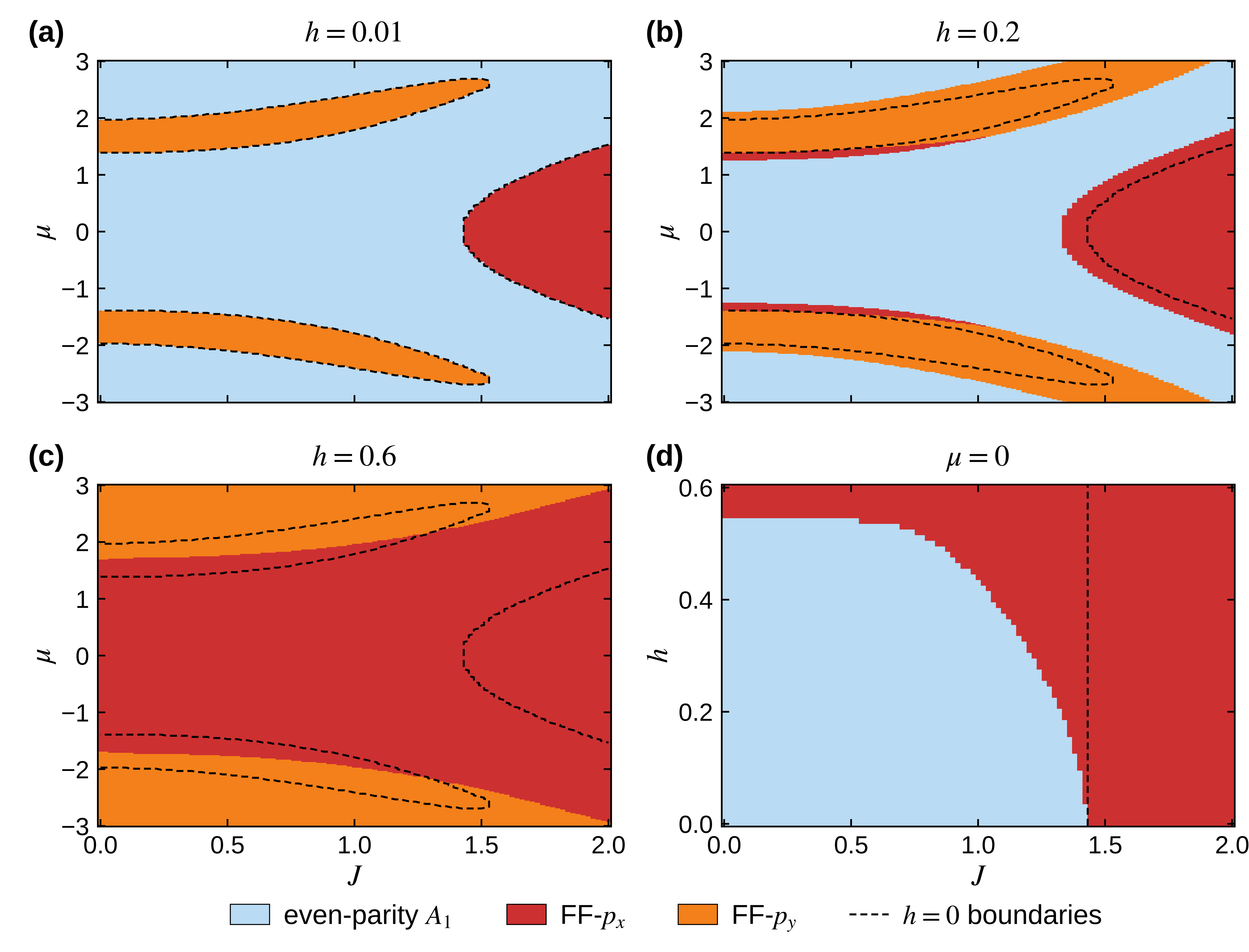}
    \caption{\textbf{Stabilization of the FF states by the Zeeman field.} (a)--(c) Phase diagrams of the superconducting ground state as a function of $J$ and $\mu$ for increasing Zeeman field: (a) $h=0.01$, (b) $h=0.2$, and (c) $h=0.6$. (d) Phase diagram as a function of $J$ and $h$ at $\mu=0$. In all panels, the parameters $(U,V)=(1,3)$ are used as in Fig.~\ref{fig2}(a). Color coding follows Fig.~\ref{fig2}. Dashed lines mark the corresponding $h=0$ phase boundaries of Fig.~\ref{fig2}.}
    \label{fig5}
\end{figure*}

To demonstrate the stabilization of the FF phases via the Zeeman effect, we recalculate the zero-field superconducting phase diagram of Fig.~\ref{fig2}(a) with the field $h$ turned on. At a tiny field $h=0.01$, the states occupy largely the same regions as in the zero-field case, but the energies of the FF-$p_x$ and FF-$p_y$ states become slightly lower than those of the odd-parity $A_1$ and $B_2$ states, respectively. As a result, the FF states become the stable superconducting phases in their respective regions [Fig.~\ref{fig5}(a)]. As the field increases, the FF-$p_x$ and FF-$p_y$ regions expand at the expense of the even-parity $A_1$ region, as shown for $h=0.2$ in Fig.~\ref{fig5}(b). At $h=0.6$, the entire examined $J$--$\mu$ parameter space ($J\le2$ and $|\mu|\le3$) is occupied by the FF-$p_x$ and FF-$p_y$ phases, and the even-parity $A_1$ phase no longer appears [Fig.~\ref{fig5}(c)]. The phase diagram in the $J$--$h$ plane at fixed $\mu=0$ shows the growing tendency of the FF-$p_x$ phase. The Zeeman field $h$ required to stabilize the FF-$p_x$ phase decreases as $J$ increases, dropping from approximately $0.55$ at $J=0$ to nearly $0$ at $J=1.44$ [Fig.~\ref{fig5}(d)]. For instance, taking $t=0.5$~eV, the field of $h\approx0.55$ corresponds to a magnetic field of approximately $4800$~T. This indicates that the FF-$p_x$ phase can be stabilized at progressively weaker Zeeman fields as the exchange field $J$ grows.

The dominance of the FF states over the zero pair-momentum states under the Zeeman field can be understood from how the field changes their Bloch Hamiltonians. In the presence of the Zeeman field $h$, the Bloch Hamiltonian for the zero pair-momentum states reads
\begin{equation}
H_\mathrm{BdG}(\bm{k}') = \begin{pmatrix}\tilde{\epsilon}(\bm{k}') - h & \Delta(\bm{k}') \\ \Delta^*(\bm{k}') & -\tilde{\epsilon}(\bm{k}') - h \end{pmatrix},
\end{equation}
where $\Delta(\bm{k}')=\Delta_{A_1}^\mathrm{even}(\bm{k}')$, $\Delta_{A_1}^\mathrm{odd}(\bm{k}')$, and $\Delta_{B_2}(\bm{k}')$ for the even-parity $A_1$, odd-parity $A_1$, and $B_2$ states, respectively. Notably, $h$ enters equally in the particle and hole sectors and, after the Bogoliubov transformation, shifts the quasiparticle energies uniformly. The odd-parity $A_1$ state has a positive minimum quasiparticle energy at $h=0$, so its quasiparticle energies remain positive for fields below this minimum, which corresponds to the Pauli limit~\cite{Casalbuoni2004}. Accordingly, in this weak-field regime, neither the pairing amplitude nor the ground-state energy changes. Once $h$ exceeds the Pauli limit, the occupation of negative-energy quasiparticle states in the superconducting ground state reduces both the pairing amplitude and the energy gain of the odd-parity $A_1$ state~\cite{Hong2025}. By contrast, the even-parity $A_1$ and $B_2$ states have nodal points where the quasiparticle energy vanishes at $h=0$, so that negative-energy quasiparticle states are occupied around these points at any finite $h$, reducing the pairing amplitude and the energy gain. A systematic analysis of the gap structures in the quasiparticle dispersions of these states is presented in Sec.~\ref{sec:experimental_signatures}.

For the FF-$p_x$ and FF-$p_y$ states, the Zeeman field enters the Bloch Hamiltonian of each spin sector with opposite signs in the particle and hole sectors,
\begin{equation}
H^\sigma_\mathrm{BdG}(\bm{k}') = \begin{pmatrix} \tilde{\epsilon}(\bm{k}') - hs_{\sigma} & \Delta_\sigma(\bm{k}')\\ \Delta_\sigma^*(\bm{k}') & -\tilde{\epsilon}(\bm{k}') + hs_{\sigma} \end{pmatrix},
\end{equation}
where $\Delta_\sigma(\bm{k}')=\Delta_{p_x;\sigma}^{\mathrm{FF}} \sin{k_x'}$ for the FF-$p_x$ state and $\Delta_\sigma(\bm{k}')=\Delta_{p_y;\sigma}^{\mathrm{FF}} \sin{k_y'}$ for the FF-$p_y$ state. The field now acts as a spin-dependent shift of the chemical potential by $+hs_{\sigma}$, which changes the size of each spin-resolved FS and thereby alters the pairing amplitude and lowers the ground-state energy even at small $h$. This is analogous to the Pauli paramagnetic response of a normal metal~\cite{RevModPhys.63.239}, and we therefore describe the energy lowering to leading order in $h$ by
\begin{equation} \label{eq:pauli_gain}
F^{\mathrm{FF}}_{\alpha}(h) = F^{\mathrm{FF}}_{\alpha}(0) - \frac{1}{2} \chi_{N} h^2 + \mathcal{O}(h^4),
\end{equation}
where $F^{\mathrm{FF}}_{\alpha}(0)$ is the ground-state energy at zero field and $\chi_{N}$ is the Pauli paramagnetic spin susceptibility defined from the normal-state ground-state energy $F^{\mathrm{N}}(h)$,
\begin{equation} \label{eq:chi_N} 
\chi_{N} = -\frac{\partial^{2}F^{\mathrm{N}}}{\partial h^{2}}\bigg|_{h=0} = \sum_{\bm{k},\sigma}\delta\big(\epsilon_{\sigma}(\bm{k})\big)\Big|_{h=0} = 2N(0),
\end{equation}
with $N(0)=\sum_{\bm{k}}\delta\big(\epsilon_{0}(\bm{k})\big)$ the density of states per spin at the Fermi level. The identification of the susceptibility of the FF states with the Pauli paramagnetic spin susceptibility holds up to corrections of order $(\Delta/E_{\mathrm{F}})^{2}$~\cite{RevModPhys.63.239}. 

Taken together, the ground-state energy of the odd-parity $A_1$ state remains unchanged in the weak-field regime below the Pauli limit and increases above it, while those of the even-parity $A_1$ and $B_2$ states increase already at weak fields. By contrast, the field lowers the ground-state energy of the FF states at any finite $h$ via the Pauli paramagnetic spin susceptibility. The energy gain of the FF states therefore exceeds that of the zero pair-momentum states at any finite $h$, so that the FF states replace the odd-parity $A_1$ and $B_2$ states and expand further at the expense of the even-parity $A_1$ state as $h$ increases [Fig.~\ref{fig5}].

We emphasize the crucial role of the spin-dependent FS shift generated by the $p$-wave exchange field. The shift makes the FF states degenerate with the zero pair-momentum states at $h=0$, and owing to this degeneracy, the FF states are favored over the zero pair-momentum states at any finite $h$. This is in contrast to conventional FF states, where the Zeeman splitting of the Fermi surfaces must be sufficiently large, comparable to the pairing gap, for a finite-momentum pair to access a larger portion of the FS as a pairing channel than a zero-momentum pair~\cite{Fulde1964, Larkin1965}. Therefore, $p$-wave magnets can serve as a platform for realizing the FF states through an intrinsic mechanism, requiring only an arbitrarily weak field.

\section{Experimental signatures} \label{sec:experimental_signatures}

\begin{figure*}
    \centering
     \includegraphics[width=.9\textwidth]{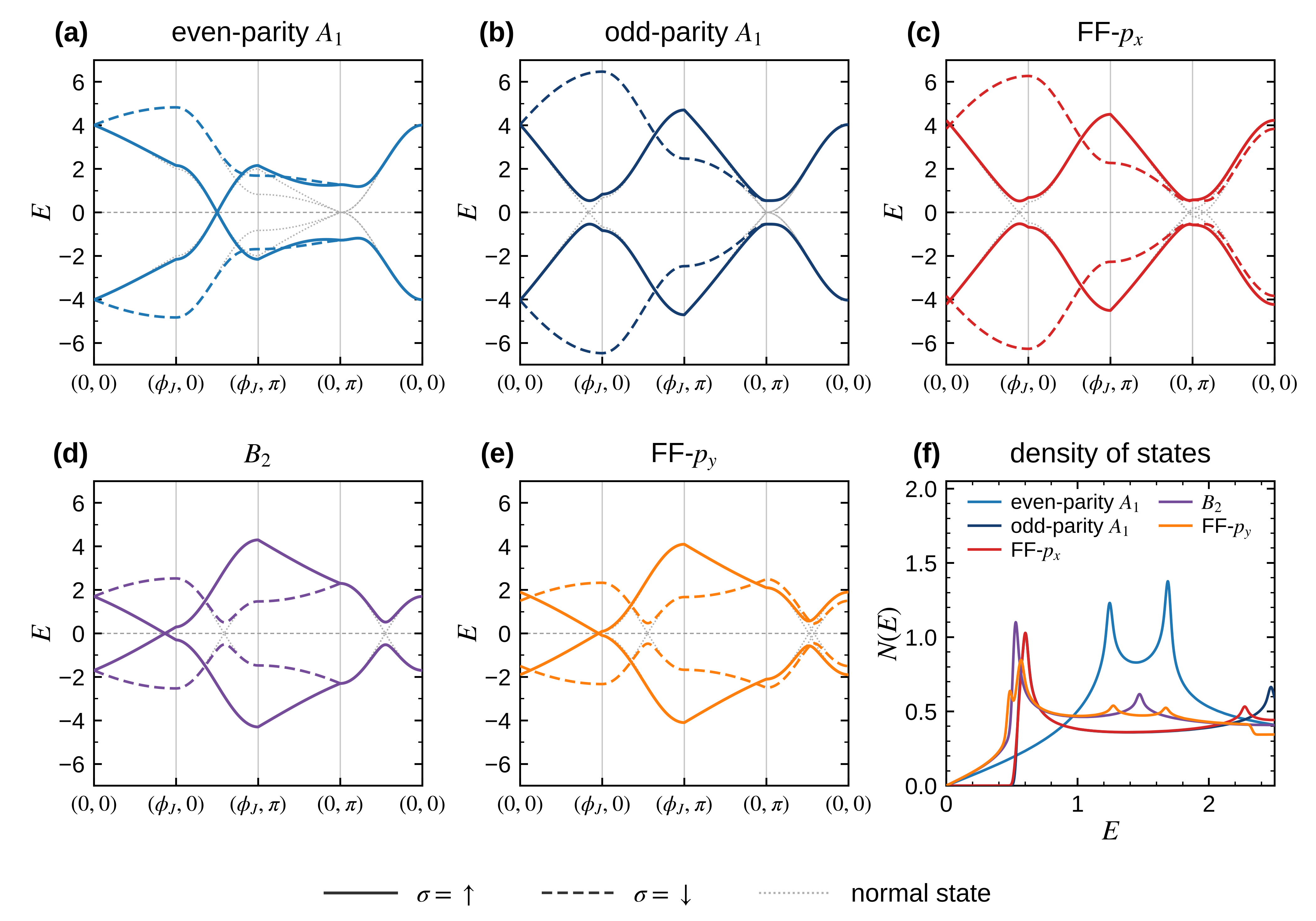}
    \caption{\textbf{Experimental signatures.} (a)--(e) Quasiparticle dispersions of (a) the even-parity $A_1$, (b) the odd-parity $A_1$, (c) the FF-$p_x$, (d) the $B_2$, and (e) the FF-$p_y$ states along the square loop of $(k_x,k_y)$, $(0,0)\to(\phi_J,0)\to(\phi_J,\pi)\to(0,\pi)\to(0,0)$, whose four segments are drawn with equal length. Solid and dashed curves correspond to $\sigma=\uparrow$ and $\downarrow$, respectively, and dotted gray curves show the normal-state dispersions $\pm\epsilon_\sigma(\bm{k})$. (f) Density of states $N(E)$ of the states in (a)--(e), computed from their quasiparticle energies with the same parameters as in the corresponding panels, summed over both spins on a $1600\times1600$ momentum grid and broadened by a Gaussian of width $0.015$. The parameters $(\mu,J,h)$ are (a) $(0,1,0)$, (b) $(0,2,0)$, (c) $(0,2,0.2)$, (d) $(-2.3,1,0)$, and (e) $(-2.3,1,0.2)$, with $(U,V)=(1,3)$ in all panels.}
    \label{fig6}
\end{figure*}

We investigate the quasiparticle dispersions as a potential experimental probe for distinguishing the different superconducting states. We focus on $(U,V)=(1,3)$ as in Fig.~\ref{fig2}(a) and take $(\mu,J,h)=(0,1,0)$, $(0,2,0)$, $(0,2,0.2)$, $(-2.3,1,0)$, and $(-2.3,1,0.2)$ for the even-parity $A_1$, odd-parity $A_1$, FF-$p_x$, $B_2$, and FF-$p_y$ states, respectively, so that each state is the stabilized ground state at the chosen parameters [Figs.~\ref{fig6}(a)--(e)]. The dispersions of the even-parity and odd-parity $A_1$ states are plotted using Eq.~\eqref{eq:E_qp_A1}, that of the $B_2$ state using Eq.~\eqref{eq:E_qp_B2}, and those of the FF-$p_x$ and FF-$p_y$ states using Eq.~\eqref{eq:E_qp_FF_shifted}.

The even-parity $A_1$ dispersion displays nodal points at $(k_x,k_y)=(\pm\pi/4,\pm\pi/2)$ for each spin, of which only the spin-up node at $(k_x,k_y)=(\pi/4,\pi/2)$ is shown in Fig.~\ref{fig6}(a). The presence of these nodal points can be understood from the condition that the normal-state dispersion and the gap function vanish simultaneously, namely $\tilde{\epsilon}(\bm{k}')=-2t_1\cos{k_x'}-2t\cos{k_y'}-\mu=0$ and $\Delta_{A_{1}}^\mathrm{even}(\bm{k}')=\Delta_s+\Delta_e\cos{k_x'}+\Delta_{s_y}\cos{k_y'}=0$. Each condition defines lines in the $k_x'$--$k_y'$ plane, and the quasiparticle energy vanishes only at their intersections. For a $d$-wave-like pattern with $\Delta_e\approx-\Delta_{s_y}$ and $\Delta_s\approx0$ [for example, the $J<1.44$ region in Fig.~\ref{fig3}(a)], the condition $\Delta_{A_{1}}^\mathrm{even}(\bm{k}')=0$ reduces to the two diagonals $k_y'=\pm k_x'$. These diagonals always cross the two segments of the contour $\tilde{\epsilon}(\bm{k}')=0$ at four points related by the mirror symmetries $k_x'\to-k_x'$ and $k_y'\to-k_y'$, giving rise to four nodal points for each spin branch of the quasiparticle energies and hence eight in total. When the $A_1$ solution deviates substantially from the $d$-wave-like pattern [for example, the $|\mu|>1.53$ region in Fig.~\ref{fig3}(b)], the two lines may in principle intersect at only two points or not at all. We find that the self-consistent values of $\Delta_s$, $\Delta_e$, and $\Delta_{s_y}$ remain in the range where the two lines intersect at four points over most of the region where the even-parity $A_1$ state is the ground state, including the even-parity $A_1$ region of the $\mu$-cut at $J=2$ in Fig.~\ref{fig3}(b), ensuring eight nodal points in total in the Brillouin zone. Only at large $|\mu|$ for small $J$, beyond $|\mu|\simeq2.0$ at $J\simeq0$, where $\Delta_e$ and $\Delta_{s_y}$ take the same sign, do the two lines no longer intersect, so that the state is fully gapped there, while a narrow band along the outer edge for $J\gtrsim1.1$ hosts only four nodal points in total (see Supplementary Note~4 for details).

By contrast, the odd-parity $A_1$ state does not possess nodal points [Fig.~\ref{fig6}(b)]. This behavior can be understood by noting that the condition $\Delta_{A_1}^\mathrm{odd}(\bm{k}')=\Delta_o\sin{k_x'}=0$ defines the straight lines $k_x'=0$ and $k_x'=\pi$. Near $\mu=0$ and $J\geq1.4$, where this state is stabilized, the exchange field elongates the two segments of the contour $\tilde{\epsilon}(\bm{k}')=0$ along $k_y'$, so that they run parallel to these lines and avoid them [Fig.~\ref{fig4}(c)]. For sufficiently small $J$ and large $|\mu|$, the contour can in principle cross these lines at two points for each spin branch, giving four nodal points in total [Fig.~\ref{fig4}(g)], but the corresponding parameters lie outside the stabilized region in the $J$--$\mu$ plane (see Supplementary Note~4). The same reasoning applies to the FF-$p_x$ state at $h=0$, whose dispersion reduces to that of the odd-parity $A_1$ state. At finite $h$, a sufficiently large field can in principle shift the spin-dependent chemical potential $\mu+hs_{\sigma}$ into the range where the two lines cross, while the FF-$p_x$ state remains stable. Such fields are, however, beyond the physically relevant range. For instance, taking $t=0.5$~eV, we find that the nodal points start to appear only for $h\gtrsim0.08$ in a limited region of $(J,\mu)$, which corresponds to a magnetic field of approximately $700$~T. The FF-$p_x$ state is thus fully gapped in practice, similarly to the odd-parity $A_1$ state [Fig.~\ref{fig6}(c)]. The two states can nevertheless be distinguished by the field-induced spin splitting of the dispersion along the $(0,0)$--$(0,\pi)$ line [Figs.~\ref{fig6}(b) and \ref{fig6}(c)]. Along this line the normal-state dispersion is spin degenerate, because the spin splitting $2Js_{\sigma}\sin k_x$ vanishes at $k_x=0$. The odd-parity $A_1$ dispersion therefore remains spin degenerate there, whereas the field shifts the two spin branches of the FF-$p_x$ dispersion in opposite directions and splits them.

The dispersion of the $B_2$ state has nodal points at $(k_x,k_y)\simeq(\pm0.68,0)$ for each spin, of which only the spin-up node at $(k_x,k_y)\simeq(0.68,0)$ is shown in Fig.~\ref{fig6}(d). Near $\mu=0$, the two segments of the contour cross the gap zeros at $k_y'=0$ and $\pi$ at four points per spin branch for $J>0$ (two for $J=0$), giving eight nodal points in total, whereas for sufficiently large $|\mu|$ they cross at two points per spin branch, giving four in total [Figs.~\ref{fig4}(d) and \ref{fig4}(h)]. In contrast to the odd-parity $A_1$ state, these nodal points are unavoidable. The gap zeros of the $B_2$ state run perpendicular to the elongation of the contour along $k_y'$, so the contour necessarily cuts through them, whereas those of the odd-parity $A_1$ state run parallel to it. Since the FF-$p_y$ state shares this nodal-point structure, with $\mu$ replaced by the spin-dependent chemical potential $\mu+hs_{\sigma}$ [Fig.~\ref{fig6}(e)], the nodal points alone cannot tell it apart from the $B_2$ state. As with the odd-parity $A_1$ and FF-$p_x$ pair, the two can instead be identified from the dispersion along the $(0,0)$--$(0,\pi)$ line, where the exchange splitting vanishes [Figs.~\ref{fig6}(d) and \ref{fig6}(e)]. Here the two spin branches of the $B_2$ dispersion coincide, while those of the FF-$p_y$ dispersion are separated by the Zeeman field.

We finally examine the density of states $N(E)$ computed from the quasiparticle energies of the states in Figs.~\ref{fig6}(a)--(e) [Fig.~\ref{fig6}(f)]. The odd-parity $A_1$ and FF-$p_x$ states show hard gaps at low energy, reflecting their fully gapped dispersions, whereas the even-parity $A_1$, $B_2$, and FF-$p_y$ states retain finite spectral weight down to zero energy because of their nodal points. Examining the existence of a hard gap therefore allows the odd-parity $A_1$ and FF-$p_x$ states to be distinguished from the even-parity $A_1$, $B_2$, and FF-$p_y$ states. Within each of these groups, the field-induced splitting pattern provides a further distinction. The odd-parity $A_1$ and FF-$p_x$ states exhibit different satellite peaks in the high-energy region $2\lesssim E\lesssim2.7$, which can be attributed to the Zeeman splitting of the FF-$p_x$ dispersion. The $B_2$ and FF-$p_y$ states, by contrast, display distinct features at lower energies, since the nodal points of the FF-$p_y$ state are displaced from those of the $B_2$ state in a spin-dependent manner by the field. Measuring the evolution of $N(E)$ under an applied field could therefore potentially discriminate among all five states.

\section{Discussion}

In this study, we discovered that the FS displacement induced by the $p$-wave exchange field provides a mechanism for realizing equal-spin, FF-type pairing states. In contrast to the classical FF states~\cite{Fulde1964}, the pairing found here incorporates the entire FS owing to the equal-energy condition, maximizing the pairing amplitude. These FF states become degenerate with the usual zero-momentum pairing states through the FS geometry, which is tuned jointly by the $p$-wave exchange field and the chemical potential, and become the leading pairing instability under arbitrarily small Zeeman fields. These findings pave the way toward realizing the long-sought FF states through an intrinsic mechanism, rather than through the extrinsic mechanisms used in previous approaches, such as strong external magnetic fields or proximity coupling to magnetic materials.

For experimental realization, we note the recent discovery of a $p$-wave magnet induced by a commensurate spin helix~\cite{Yamada2025}, which gives rise to the required $p$-wave exchange field. Alternatively, one could consider orbital order~\cite{Zhuang2026}, or induce the required odd-parity exchange field via light irradiation of conventional collinear antiferromagnets through Floquet engineering~\cite{9346-9jpf,7ywb-ml2q}; see Ref.~\cite{Zeng2026} for a systematic symmetry classification of $p$-wave magnetism. The formation of the FF-$p_x$ state could then be detected by tracking the unconventional, spin-split nodal-point structure of the BdG quasiparticle bands identified here, via scanning tunneling spectroscopy and, in particular, Bogoliubov quasiparticle interference imaging, which directly resolves the momentum dependence of the excitation gap~\cite{Hanaguri2010,Allan2012}.

It remains an important task for future work to extend the current effective-model approach to more realistic model setups. Specifically, one can employ two-sublattice models~\cite{Khodas2026, Kim2026a} that capture the underlying spin texture responsible for the $p$-wave exchange field, or incorporate spin-orbit coupling; the latter might affect the subtle energetic competition between the FF and non-FF states observed here, thereby controlling their instability. One might also consider higher-order exchange fields ($d$-, $f$-, $g$-wave types, etc.)~\cite{Smejkal2022b} to explore their realization. For this direction, however, we point out that these higher-order exchange fields are not expected to induce an FS displacement, a crucial distinction that limits the direct applicability of the matched-momentum mechanism. In this regard, we argue that the $p$-wave exchange field plays a distinctive role in stabilizing finite-momentum pairing states.

\section*{Methods}

\subsection*{Gap equations of $A_1$ states}

Evaluating the anomalous averages in Eq.~\eqref{eq:gap_eq_A1} at zero temperature casts the gap equations in the explicit form~\cite{Kim2026a}
\begin{equation} \label{eq:gap_eq_A1_explicit}
\bm{\Delta}_{A_1}=\mathcal{M}\,\bm{\Delta}_{A_1},
\end{equation}
where $\bm{\Delta}_{A_1}\equiv(\Delta_{s},\Delta_{s_x},\Delta_{s_y},
\Delta_{p_x})^{\mathrm{T}}$ and
\begin{equation} \label{eq:M_A1}
\mathcal{M}=
\begin{pmatrix}
U\mathcal{I}_{s,s} & U\mathcal{I}_{s,s_x} & U\mathcal{I}_{s,s_y} & U\mathcal{I}_{s,p_x}\\
2V\mathcal{I}_{s_x,s} & 2V\mathcal{I}_{s_x,s_x} & 2V\mathcal{I}_{s_x,s_y} & 2V\mathcal{I}_{s_x,p_x}\\
2V\mathcal{I}_{s_y,s} & 2V\mathcal{I}_{s_y,s_x} & 2V\mathcal{I}_{s_y,s_y} & 2V\mathcal{I}_{s_y,p_x}\\
2V\mathcal{I}_{p_x,s} & 2V\mathcal{I}_{p_x,s_x} & 2V\mathcal{I}_{p_x,s_y} & 2V\mathcal{I}_{p_x,p_x}
\end{pmatrix},
\end{equation}
with the overlap integrals
\begin{equation} \label{eq:overlap_A1}
\mathcal{I}_{\alpha,\beta}=\frac{1}{N}\sum_{\bm{k}}
\frac{g_{\alpha}(\bm{k})\,g_{\beta}(\bm{k})\Theta\big(E^{A_1}_{\up}(\bm{k})\big)}{2\sqrt{\epsilon_{0}(\bm{k})^{2}+\bigl|\Delta_{A_1}(\bm{k})\bigr|^{2}}} ,
\end{equation}
for $\alpha,\beta\in\{s,s_x,s_y,p_x\}$. Here $\epsilon_{0}(\bm{k})=\frac{1}{2}\big[\epsilon_{\up}(\bm{k})+\epsilon_{\dn}(-\bm{k})\big]$ denotes the symmetric combination of the dispersions of the two paired electrons, in which the Zeeman term cancels,
\begin{equation} \label{eq:E_ns_3}
    \epsilon_{0}(\bm{k}) = -2t_1 \cos\left(k_x + \phi_J\right) -2t\cos k_y  - \mu,
\end{equation}
and $E^{A_1}_{\sigma}(\bm{k})$ denotes the quasiparticle energy
\begin{equation} \label{eq:E_qp_A1}
E^{A_1}_{\sigma}(\bm{k})=\sqrt{\epsilon_{0}(s_{\sigma}\bm{k})^{2}+\bigl|\Delta_{A_1}(s_{\sigma}\bm{k})\bigr|^{2}} - hs_{\sigma}.
\end{equation}
The ground-state energy is
\begin{equation} \label{eq:Fcond_A1}
\begin{aligned}
F_{A_1}={}& \sum_{\bm{k}}\Bigl[\epsilon_{\dn}(\bm{k}) - E^{A_1}_{\dn}(\bm{k})\Bigr] +\frac{N}{2V}\!\!\sum_{\alpha=s_x,s_y,p_x}\!\!|\Delta_\alpha|^2 \\
& + \frac{N}{U}|\Delta_{s}|^2 + \sum_{\bm{k}}\Theta\big(-E^{A_1}_{\up}(\bm{k})\big)E^{A_1}_{\up}(\bm{k}). 
\end{aligned}
\end{equation}
The derivations of these equations are given in Supplementary Note~5.

\subsection*{Gap equations of $B_2$ states}

Evaluating the anomalous average in Eq.~\eqref{eq:gap_eq_B2} at zero temperature casts the $B_2$ gap equation in the explicit form
\begin{equation} \label{eq:gap_eq_B2_explicit}
\Delta_{p_y}=\frac{V}{N}\sum_{\bm{k}}
\frac{\Delta_{p_y}g_{p_y}(\bm{k})^2\Theta\big(E^{B_2}_{\up}(\bm{k})\big)}{\sqrt{\epsilon_{0}(\bm{k})^{2}
+\bigl|\Delta_{B_2}(\bm{k})\bigr|^{2}}},
\end{equation}
where the quasiparticle energies of the two paired modes are
\begin{equation}\label{eq:E_qp_B2}
E^{B_2}_{\sigma}(\bm{k}) = \sqrt{\epsilon_0(s_{\sigma}\bm{k})^2+
|\Delta_{B_2}(s_{\sigma}\bm{k})|^2}-hs_{\sigma}.
\end{equation}
The ground-state energy is
\begin{equation} \label{eq:Fcond_B2}
\begin{aligned}
F_{B_2}={}& \sum_{\bm{k}}\Bigl[\epsilon_{\dn}(\bm{k}) - E^{B_2}_{\dn}(\bm{k})\Bigr] +\frac{N}{2V}|\Delta_{p_y}|^2 \\
& + \sum_{\bm{k}}\Theta\big(-E^{B_2}_{\up}(\bm{k})\big)E^{B_2}_{\up}(\bm{k}).
\end{aligned}
\end{equation}
The derivations of these equations are given in Supplementary Note~6.

\subsection*{Gap equations of FF states}

Evaluating the anomalous average in Eq.~\eqref{eq:gap_eq_FF} at zero temperature brings this to the standard form
\begin{equation}\label{eq:gap_eq_FF_explicit}
\Delta_{\alpha;\sigma}^{\mathrm{FF}}=\frac{V}{N}\sum_{\bm{k}}\frac{\Delta_{\alpha;\sigma}^{\mathrm{FF}}|g_\alpha(\bm{k})|^2}{\sqrt{\tilde{\epsilon}_\sigma(\bm{k})^2+\bigl|\Delta_{\alpha;\sigma}^{\mathrm{FF}} g_{\alpha}(\bm{k})\bigr|^2}},
\end{equation}
for $\alpha=p_x,p_y$. The ground-state energy is
\begin{equation} \label{eq:Fcond_FF}
F^{\mathrm{FF}}_{\alpha} = \frac{1}{2}\sum_{\bm{k},\sigma}\Bigl[\tilde{\epsilon}_\sigma(\bm{k})-\tilde{E}_{\alpha;\sigma}^{\mathrm{FF}} (\bm{k}) \Bigr] + \frac{N}{4V}\sum_{\sigma}\bigl|\Delta_{\alpha;\sigma}^{\mathrm{FF}}\bigr|^2,
\end{equation} 
where the quasiparticle energy is
\begin{equation}\label{eq:E_qp_FF}
\tilde{E}_{\alpha;\sigma}^{\mathrm{FF}} (\bm{k})=\sqrt{\tilde{\epsilon}_\sigma(\bm{k})^2+\bigl|\Delta_{\alpha;\sigma}^{\mathrm{FF}} g_{\alpha}(\bm{k})\bigr|^2}.
\end{equation}
The derivations of these equations are provided in Supplementary Note~7. 

To plot the quasiparticle dispersions for the FF-$p_x$ and FF-$p_y$ states [Figs.~\ref{fig6}(c) and (e)], we used
\begin{equation} \label{eq:E_qp_FF_shifted}
    E_{\alpha;\sigma}^{\mathrm{FF}} (\bm{k}) = \sqrt{\epsilon_{\sigma}(\bm{k})^{2} + \left|\Delta_{\alpha;\sigma}^{\mathrm{FF}}\, g_{\alpha}\!\left(\bm{k}-\frac{\bm{Q}_\sigma}{2}\right)\right|^{2}},
\end{equation}
which follows from Eq.~\eqref{eq:E_qp_FF} by the substitution $\bm{k}\rightarrow \bm{k} - \bm{Q}_\sigma/2$ for each spin $\sigma$, so that the energies of both spins are expressed as functions of the same physical momentum.

\subsection*{Solving gap equations}

The gap equations [Eqs.~\eqref{eq:gap_eq_A1_explicit}, \eqref{eq:gap_eq_B2_explicit}, and \eqref{eq:gap_eq_FF_explicit}] were solved iteratively. All momentum sums were evaluated in the shifted coordinates $\bm{k}'$ of Eq.~\eqref{eq:shifted_momentum} on a grid of $300\times300$ points, so that all four pairing states are described by the same dispersion $-2t_1\cos k_x'-2t\cos k_y'-\mu$ on the same grid. At each step, the right-hand side of the gap equation was evaluated with the current pairing amplitudes and taken as the amplitudes of the next step, an update which typically does not increase the ground-state energy. The iteration was terminated when the change of every amplitude between successive steps fell below $10^{-8}$, or after 300 steps. When all amplitudes fell below $10^{-6}$, the solution was identified as the normal state. For the coupled $A_1$ gap equations, the iteration was started from four initial configurations, $(\Delta_s,\Delta_{s_x},\Delta_{s_y},\Delta_{p_x})=(1,1,1,0)$, $(0,-1,1,0)$, $(0,0,0,1)$, and $(0,-1,1,1)$, corresponding to $s$-wave-like, $d$-wave-like, odd-parity, and mixed-parity configurations, respectively, and the solution with the lowest ground-state energy was retained. The $B_2$ and FF gap equations, each involving a single amplitude, were iterated from $\Delta_{p_y}=1$ and $\Delta^{\mathrm{FF}}_{\alpha;\sigma}=1$, with the FF gap equation solved separately for each spin $\sigma$. The resulting pairing amplitudes were substituted into the ground-state energy expressions [Eqs.~\eqref{eq:Fcond_A1}, \eqref{eq:Fcond_B2}, and \eqref{eq:Fcond_FF}], and the state with the lowest energy was identified as the superconducting ground state, from which the phase diagrams in Figs.~\ref{fig2} and \ref{fig5} were constructed. Energies differing by less than $10^{-7}$ were regarded as degenerate, and the normal state was assigned unless a superconducting state lowered the energy by more than $10^{-7}$.

\begin{acknowledgements}
This research was supported by an appointment to the JRG Program at the APCTP through the Science and Technology Promotion Fund and Lottery Fund of the Korean Government, the Korean Local Governments (Gyeongsangbuk-do Province and Pohang City), and the National Research Foundation of Korea (NRF) funded by the Korean government (Ministry of Science and ICT, MSIT) (No. RS-2026-25499525).
\end{acknowledgements}

\bibliography{ref}

\end{document}